%% file: SMGT2024-v04.tex
\documentclass[
prl,
 reprint,
 amsmath,amssymb,
 aps,
]{revtex4-2}

\usepackage{graphicx}
\usepackage{dcolumn}
\usepackage{bm}
\usepackage{xcolor}
\usepackage{amsmath}
\usepackage{amssymb}
\usepackage[utf8]{inputenc}
\usepackage[T1]{fontenc}
\usepackage{mathptmx}
\usepackage{etoolbox}
\begin{document}
%\preprint{APS/123-QED}
\title{Localized States in Dipolar Bose-Einstein Condensates: To be or not to be of second order}

\author{A. B. Steinberg}
\altaffiliation{ORCID: 0000-0001-6598-9700}
\affiliation{Institut für Theoretische Physik, Universität Münster, Wilhelm-Klemm-Strasse 9, 48149 Münster, Germany}

\author{F. Maucher}
\altaffiliation{ORCID: 0000-0002-5808-3967}
\affiliation{Faculty of Mechanical Engineering;
Department of Precision and Microsystems Engineering,
Delft University of Technology, 2628 CD, Delft, The Netherlands}

\author{S. V. Gurevich}
\altaffiliation{ORCID: 0000-0002-5101-4686}

\author{U. Thiele}
\altaffiliation{ORCID: 0000-0001-7989-9271; $\quad$ Electronic mail: u.thiele@uni-muenster.de}
\homepage{http://www.uwethiele.de}
\affiliation{Institut für Theoretische Physik, Universität Münster, Wilhelm-Klemm-Strasse 9, 48149 Münster, Germany}
\affiliation{Center of Nonlinear Science (CeNoS), Universität Münster, Corrensstrasse 2, 48149 Münster, Germany}

\date{\today}

\begin{abstract}
We report the existence of localized states in dipolar Bose-Einstein condensates confined to a tubular geometry. We first perform a bifurcation analysis to track their emergence in a one-dimensional domain for numerical feasibility and find that localized states can become the ground state in suitable parameter regions. Their existence for parameters featuring a supercritical primary bifurcation shows that the latter is not sufficient to conclude that the phase transition is of second order, hence density modulations can jump rather than emerging gradually. Finally, we show that localized states also exist in a three-dimensional domain.
\end{abstract}

\maketitle

In 1938 it was found that at very low temperatures Helium can be superfluid, i.e., behave like a liquid without any viscosity or shear~\cite{Kapitza1938,ALLEN1938}. Roughly 30 years after that speculations were made whether solids, i.e. states with discrete translational symmetry, could display similar behavior~\cite{AnLi1969spjetp,Ches1969pra,Legg1970prl}. Whereas this has not yet been confirmed for Helium, Bose-Einstein condensates (BECs) emerged as an ideal platform for experiments on these so-called supersolids, which were realized in 2017 by two experimental groups using additional light fields~\cite{LMZE2017nature,LLHB2017nature}.

Dipolar BECs can self-organize into supersolids without required additional fields, stabilized by beyond mean-field contributions~\cite{LiPe2011pra,LiPe2012pra,WaSa2016pra,Hiroki:JPhysJap:2016} and first observed experimentally in~\cite{Kadau2016,SWBF2016nature}. Subsequently, dipolar BECs became a unique experimental setting for studying quantum fluctuations, pattern-formation, phase-transitions and supersolidity~\cite{Kadau2016,CPIN2019prx,BSWH2019prx,TLFC2019prl,TRLF2019nature,NBPP2019prl,GBHS2019nature,TMBF2021science,Petter:PRA:2021,IKSP2018pra,NPKP2021nature,Norcia:PRA:2022} accompanied by theoretical advances~\cite{WaSa2016pra,BWBB2016pra,Bombin:PRL:2017,Roccuzzo:PRA:2019,GBHS2019nature,ZhMP2019prl,BaBP2020ctp,BBCF2020prr,ZhPM2021pra,HSGB2021prr,Hertkorn:PRL:2021,SmBB2023pra,Zhang:Atoms:2023,Sanchez-Baena:NatCom:2023,Baena:PRR:2024,Zhang:PRR:2024}. The competition between dipolar interactions, quantum fluctuations and scattering leads to intricate nonlinear dynamics and a rich phase diagram~\cite{ZhMP2019prl,ZhPM2021pra,HSGB2021prr,Zhang:PRR:2024}. In particular, the change from a superfluid to a supersolid can proceed via a first- or second order phase transition. A second-order phase transition has been predicted to occur at a single point for a two-dimensional infinite pancake geometry with two-dimensional symmetry-breaking~\cite{ZhMP2019prl} and in an extended region for a three-dimensional tubular geometry with one-dimensional symmetry-breaking~\cite{BaBP2020ctp,BBCF2020prr,SmBB2023pra,Baena:PRR:2024}.

In the context of BECs, the order of the phase transition is usually concluded via complex time evolution and by comparing the energies of the converged states, thereby assuming that the average local density is uniform across the system. Yet, in the context of colloidal crystallization it has been pointed out that there is a subtlety when exploring the order of phase transitions in systems with particle number (or norm) conservation~\cite{TARG2013pre,TFEK2019njp,HAGK2021ijam}, in particular in the thermodynamic limit. Namely, when using an average density as control parameter, a supercritical bifurcation is not sufficient to conclude that the phase-transition is of second-order. Instead, one may consider the existence of localized states as a direct signature of the occurrence of a phase transition of first order. As we will see in more detail, the reason as to why such states can exist and even become ground states is due to the possibility of distributing the local average  density nonuniformly across the domain. In other words, states of different mean density may coexist if they have identical chemical potential and grand potential~\cite{TFEK2019njp,HAGK2021ijam}.

\begin{figure}[h]
    \centering
    \includegraphics[width=0.95\linewidth]{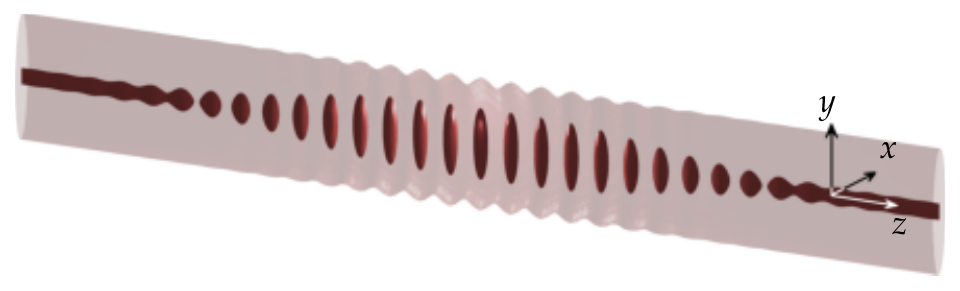}
    \caption{Example of a stable localized state in a three-dimensional dipolar BEC system. The dipoles are polarized along the $y$-direction, the BEC is unconfined in the $z$-direction (axially) and tightly confined in the transverse direction. Even without trap in $z$-direction, there exist stable localized states which feature a modulated supersolid in the central region and a perfect superfluid sufficiently far from it. This corresponds to an unequal distribution of the local average density.
    The parameters are $n=5250/\mu$m and $a_s=91a_0$. Contours are given at $1\%$ (light) and $75\%$ (dark) of the peak density. For ease of viewing the $z$-direction is compressed by a factor 7.\label{fig:3D_loc}}
\end{figure}

Localized states are common phenomena that can be observed in various pattern forming system~\cite{TlML1994prl,BuKn2006pre,LBDC2008siam,TARG2013pre,Knob2015arcmp,Knob2016ijam}. In the present BEC context~\cite{HeBD2019pra,SMGT2022cha} it refers to a self-organized density configuration where a finite patch of spatially modulated density resides in a background of uniform density.
In other words, when moving along a line through the system one first finds a perfect superfluid that then transitions via modulations of increasing amplitude into a fully modulated supersolid or droplet crystal. Figure~\ref{fig:3D_loc} shows such a localized (stationary) state in a three-dimensional dipolar BEC.

Although, in general, the existence of such localized states was conjectured for BECs with long-range interactions~\cite{HeBD2019pra,SMGT2022cha}, to our knowledge none have yet been reported for realistic fully nonlocal systems of BECs most likely due to computational limitations. This is due to the required system size and because the nonlocal nonlinearity typically is numerically expensive.

In this letter, we perform a bifurcation analysis for systems of the required large size employing models that incorporate the fully nonlocal nonlinearity and employing nonlocal extensions of the numerical bifurcation techniques provided by \texttt{pde2path}~\cite{UeWR2014cup}. In particular, we report that localized states indeed exist in systems with nonlocal or long-range interactions, namely, in dipolar BECs. Their existence implies that the underlying phase transition is of first order and is a direct consequence of the possible coexistence of two phases (i.e., modulated and uniform).

First, we completely analyze an effective one-dimensional system for reasons of numerical feasibility: We present a full bifurcation analysis including localized states and discuss a related double-tangent Maxwell construction~\cite{callen1960thermodynamics}, i.e.,
we identify two states of different phase (and mean density) that have identical grand potential per particle and chemical potential. Second, we show that localized states are also possible in three-dimensional systems.

We consider an ensemble of $N$ dipolar atoms of mass $m$ at zero temperature that interact via both, collisions and dipolar long-range interactions, characterized by $a_{\rm s}$ and $a_{\rm dd}=m\mu_0\mu_m^2/12\pi\hbar^2$,  respectively, with $\mu_m$ being the magnetic moment, and $\mu_0$ the magnetic constant. For $^{164}$Dy it is $m=163.93\,$u. The dipolar interaction is described by
\begin{align}
    U(r) &= \frac{4\pi \hbar^2a_s}{m} \delta(r) + \frac{\hbar^2}{m}\frac{3 a_{dd}}{r^3}\left(1-3\frac{y^2}{r^2}\right)\,.\label{eq:interaction-pot}
\end{align}
Here, we assume that the dipoles are oriented along the $y$-direction. External traps tightly confine the wave function $\psi$ into a tubular geometry, such that the dipoles have a side-by-side orientation along the unconfined axial direction ($z$) and are strongly confined in the polarization direction of the dipoles ($y$) and the remaining direction ($x$). Assuming equal trapping frequencies $\omega_x=\omega_y=\omega=2\pi\times 150\,$Hz, the trapping potential is $V(\bm r)=V(x,y)=\frac{1}{2}\omega^2(x^2+y^2)$. Furthermore, the wave function is normalized to the total particle number $N=\int|\psi|^2d^3r$.

The energy per particle is given by 
\begin{align}
    \frac{E}{N}&=\frac{1}{N}\int\frac{\hbar^2}{2m}|
    \nabla \psi|^2+V(\bm r)|\psi|^2\nonumber\\
    &+\frac{1}{2}|\psi|^2\int U({\bm r}-{\bm r}^\prime)|\psi({\bm r}^\prime)|^2d^3r^\prime+\frac{2}{5}\gamma_{QF}|\psi|^5d^3r \,,\label{eq:gpe}
\end{align}
with $\gamma_{QF} = \frac{\hbar^2}{m}4\sqrt{\pi} \frac{32}{3}\sqrt{a_s}^5\int_0^1\text{d}u\left(1+\frac{a_{dd}}{a_{s}}(3u^2-1)\right)^{5/2}$ \cite{LiPe2011pra}. To describe the dynamics we employ the extended Gross-Pitaevskii equation which corresponds to $i\hbar \partial_t \psi = \frac{\delta E}{\delta\psi^\star}$ with Eq.~\eqref{eq:gpe}.

We proceed by considering an effective one-dimensional system: we assume a Gaussian profile perpendicular to the axial direction and integrate out the transverse direction. More specifically, following Ref.~\cite{BBCF2020prr} we assume $\psi({\bm r},t)=\psi_{\parallel}(z)\psi_{\perp}(x,y)e^{-it\mu/\hbar}$ with chemical potential $\mu$ and $\psi_{\perp}(x,y)=\frac{1}{\sqrt{\pi}l}e^{-(\eta x^2+y^2/\eta)/2l^2}$~\footnote{In principle the parameters $l$ and $\eta$ that describe the transverse profile should be adapted when aiming at quantitative agreement. However, for localized states the optimal values of these parameters depend on position, i.e., they vary between modulated region and the uniform background (cf.~Fig.~\ref{fig:3D_loc}).}. It is useful to introduce the mean density $\bar n=N/L$, where $L$ denotes the length of the tube and the one-dimensional density $n(z)=\int|\psi|^2dxdy$, which in case of the Gaussian transverse profile gives $n(z)=|\psi_\parallel(z)|^2$. For simplicity, here we fix these parameters as $l=1.1456$ and $\eta=5.8077$ -- chosen as optimal values at the primary bifurcation ($\bar n=\bar{n}_\mathrm{c}$) for $a_{\rm s}=89a_0$. The effective model successfully predicts qualitative features, however, it is not fully quantitative  as the actual transverse profile deviates from a Gaussian~\cite{BaBP2020ctp,BBCF2020prr,SmBB2023pra}. However, it permits us to perform a bifurcation analysis and facilitates a complete understanding of the underlying physics. The results provide a proof of principle that localized states exist and allow for a discussion of their general properties. Below, also fully three-dimensional localized states will be computed via complex time evolution showing that the dimensional reduction is qualitatively faithful. 

\begin{figure}[h]
    \centering
    \includegraphics[width=0.99\linewidth]{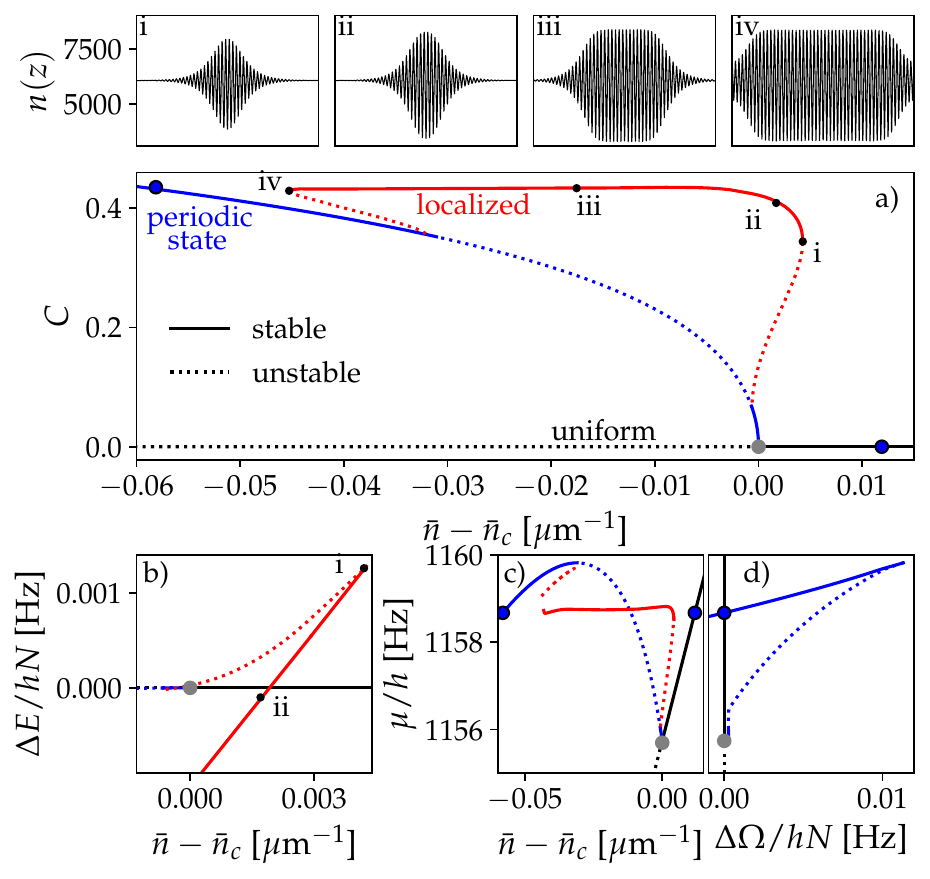}
    \caption{Localized states in effective one-dimensional dipolar BEC model. Panels (a), (b), and (c) show the bifurcation diagram in terms of the contrast $C$, the mean energy per particle $\Delta E/N$, and the chemical potential $\mu$, respectively, as function of the mean density $\bar n - \bar n_\mathrm{c}$ including branches of uniform perfect superfluid states (black), of periodic states (blue) and of localized states (red). Solid [dotted] lines indicate linearly stable [unstable] states. Profiles $n(z)$ of the localized states marked i to iv in (a) are depicted in the top row. The presence of localized states indicates that the superfluid-supersolid phase transition is of first order: Panel (b) indicates  that the energy of the localized state is lower (e.g., state ii) than the energy of the superfluid even for densities larger than the critical one ($ \bar n_\mathrm{c}$, gray circle). Panel (c) clearly shows a region of constant $\mu$ indicating the localized states represent the coexistence of different states, as further evidenced by panel (d) by the crossing of two stable branches in the $(\Delta\Omega/N,\mu)$-plane, where $\Delta\Omega/N$ is the grand potential per particle. The corresponding states are marked by filled circles in (a). The remaining parameters are $a_s = 98 a_0$, $l=1.1456$ and $\eta=5.8077$.}
    \label{fig:as89_c}
\end{figure}

Performing an integration over the transverse dimensions yields an effective one-dimensional interaction of the form 
$\hat{U}(k) = \frac{\hbar^2}{m}\frac{2}{l^2}\left(a_s+a_{dd}\left[\frac{3(Qe^Q\text{Ei}(-Q))}{\eta+1}-1\right]\right)$, 
where $Q=(\sqrt{\eta}k^2l^2)/2$ and Ei$(x)$ is the exponential integral function. Dimensional reduction and expansion of the quantum fluctuations~\cite{BaBP2020ctp,BBCF2020prr} amounts to 
$\gamma_{QF} \approx \frac{\hbar^2}{m}\frac{256}{15\pi}\frac{a_s^{5/2}}{l^3}\left(1+\frac{3a_{dd}^2}{2a_s^2}\right)$. We apply numerical path continuation~\cite{UeWR2014cup, EGUW2019springer} to determine a bifurcation diagram, i.e., a diagram that shows branches of metastable and unstable states in addition to the stable states usually found through complex time evolution which replaces $t\rightarrow-it$ and renormalizes the wave function after each propagation step. Fig.~\ref{fig:as89_c} provides such a diagram for $a_{\rm s}=89a_0$ as a function of the mean density $\bar n$. The possible stationary states can be uniform (whereby we mean  uniform with respect to the $z$-direction), periodically modulated, i.e., with discrete translational symmetry, or localized. An upper bound for the associated superfluid fraction can be found using the Leggett estimator~\cite{Legg1970prl,Leggett1998}, which in case of the uniform state would amount to a perfect superfluid and for a modulated density decreases as function of the modulation amplitude~\footnote{Where necessary path continuation results are refined using complex time evolution. In case of meta- and unstable state data is extrapolated from neighboring results.}.

Figure~\ref{fig:as89_c}~(a) presents the bifurcation diagram in terms of the contrast $C = \frac{|\psi_{\parallel}|^2_\text{max}-|\psi_{\parallel}|^2_\text{min}}{|\psi_{\parallel}|^2_\text{max}+|\psi_{\parallel}|^2_\text{min}}$ as a function of the one-dimensional mean density $\bar{n}-\bar{n}_\mathrm{c}$ where $\bar{n}_{\rm c}=6060.49/\mu$m. For densities larger than $\bar{n}_{\rm c}$ the uniform density state is linearly stable due to the pattern-inhibiting quantum fluctuations (black solid line). The primary bifurcation at $\bar{n}=\bar{n}_\mathrm{c}$ (gray circle) corresponds to a symmetry-breaking supercritical pitchfork bifurcation. There, the superfluid state loses stability and a stable supersolid state featuring periodic density modulations emerges (solid blue line).

As discussed in Refs.~\cite{BBCF2020prr,SmBB2023pra}, in the tubular geometry for assumed uniform local mean density, this phase transition can be either of first or second order, depending on the parameter region. For a very small system no localized states (red line) can develop. In consequence, Fig.~\ref{fig:as89_c}~(a) indicates a second-order transition. However, this does not apply for sufficiently large systems and, indeed, the thermodynamic limit. Then, also localized states do exist (red line, example profiles i to iv in top row of Fig.~\ref{fig:as89_c}), and even represent the ground state. The branch of unstable localized states that represents critical nuclei bifurcates subcritically from the branch of periodic states (that also become unstable), continues first towards higher densities while the contrast $C$ increases. Then, at a saddle-node bifurcation (state i) that represents a point of maximal density, the branch stabilizes and turns back towards lower $\bar n$. Shortly after, $C$ reaches a plateau that it keeps until another limiting saddle-node bifurcation at lower $\bar n$ (state iv). As a result, there exist extended bistable regions -- one for $\bar n>\bar{n}_c$ between uniform and localized state and another one at lower $\bar n$ between localized and regular periodic state. Ultimately, after another unstable part, the branch of localized states ends on the branch of periodic states thereby stabilizing them. Beside the described bistable ranges, interestingly, there exists a rather extended density range where the localized states are the only stable ones.

Consider now Fig.~\ref{fig:as89_c}~(b), which gives the energy difference $\Delta E=E-E_0$ between the various states and the perfectly uniform superfluid. It is interesting that state ii is the energetically most favored state even at a density that is larger than $\bar n_\mathrm{c}$ where the primary bifurcation occurs (gray point). Hence, upon decreasing the density from a large value and performing complex time evolution we first jump to state ii before reaching the primary bifurcation. This is accompanied by a jump in contrast as shown in Fig.~\ref{fig:as89_c}~(a) and, therefore, corresponds to a first-order phase transition.

Figure~\ref{fig:as89_c}~(c) shows the chemical potential as a function of the density. The chemical potential remains approximately constant for localized states in the domain where they are stable. This is characteristic for localized states, that roughly follow the line of the double-tangent Maxwell construction~\cite{callen1960thermodynamics} even for finite systems, i.e., outside the thermodynamic limit~\cite{TFEK2019njp}.

This implies that the chemical potential and pressure (equal to the negative grand potential per particle) are identical for coexisting phases. To better understand the origin of localized states we furthermore consider the grand potential per particle $\frac{\Omega}{N} = \frac{E}{N}-\mu $.
Usually, $\Omega$ is employed in BECs when considering finite temperature effects~\cite{Griffin:PRB:1996,Ronen:PRA:2007,Sanchez-Baena:NatCom:2023}. However, here we consider the case $T=0$.
  
To identify regions of coexisting phases, one depicts all branches in the $(\mu,\Omega/N)$-plane [Fig.~\ref{fig:as89_c}~(d)], to check for crossings of the superfluid and periodic state branches. Such a crossing indeed exists and the coexisting states are marked by blue filled circles in Figs.~\ref{fig:as89_c}~(a,\,c,\,d). The interval of densities between these points is where localized states can exist. The finite numerical domain results in a slightly smaller existence region. 

\begin{figure}[h]
    \centering
    \includegraphics[width=0.99\linewidth]{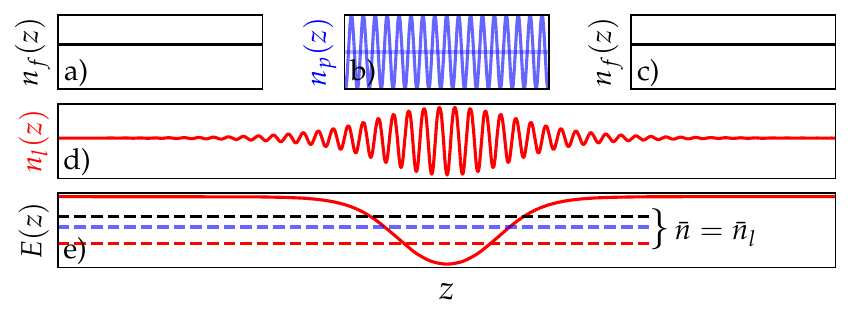}
    \caption{Three ground states (a)-(c) that are isolated with respect to each other. The mean density of the modulated state ((b), blue) features a slightly smaller mean density, as indicated by the gray line. Removing the separating walls leads to a flux of density up to an equilibrium (d), in this case (corresponding to state iii). The ground state actually energetically prefers to retain a low-density modulated region in the center and an uniform state far from center as this is still energetically favorable as opposed to the uniform (black) and periodic states (blue) ((e), dashed lines). The energy differences are exaggerated to dramatize the situation. The local energy of the localized state is given as a red solid line.}
    \label{fig:thought_experiment}
\end{figure}

We would like to gain further insight as to why such states emerge. In a thought experiment visualized in Fig.~\ref{fig:thought_experiment} we imagine two boxes that each contains a different ground state, namely, a uniform and a periodic state, (see Fig.~\ref{fig:thought_experiment}~(a)-(c)). The chosen states can have different mean densities. Bringing the boxes together and removing the boundaries between them, we let the system relax to a new ground state, in a simulation via complex time-evolution. This leads to a flux such that finally $\mu$ becomes globally uniform. Depending on the parameters, the dynamics could either crystallize the entire system, dampen out all modulations, or only smoothen the transition between the two states, i.e., get pinned in a localized state that has the local mean density distributed in a nonuniform fashion, (cf. Fig.~\ref{fig:thought_experiment}~(d)). In fact, such a configuration with nonuniform local mean density can also be energetically preferable. Consider the resulting localized state in Fig.~\ref{fig:thought_experiment}: At the center of the system the state has lower local mean density and energy per particle. In the outer parts mass and energy are higher. However, their combination gives a total mean energy (red horizontal line in Fig.~\ref{fig:thought_experiment}~(e)) that can be lower than the ones of the superfluid (black) and the periodic (blue) states.

\begin{figure}[h]
    \centering
    \includegraphics[width=0.99\linewidth]{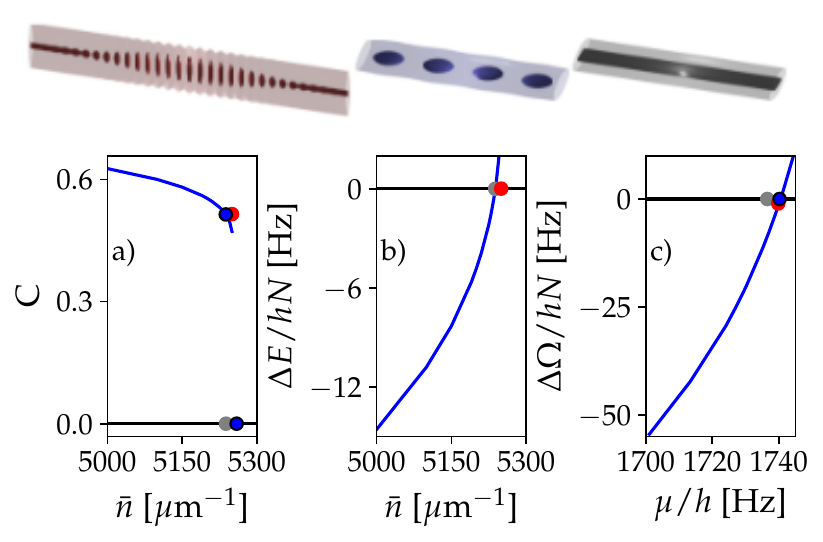}
    \caption{States for a system of $\bar{n} =5290/\mu$m. Top row: Localized state (left, red), periodic state (middle, blue) and superfluid (right, black) with contours at $1\%$ (light) and $75\%$ (dark) of the peak density. For ease of viewing the $z$ direction is compressed by a factor 7 [2] for the localized state [periodic and superfluid states].
    (a) displays the contrast as a function of the mean density for the state with discrete translational symmetry (blue), the uniform state (black), a localized state (red dot) and the points of thermodynamic coexistence (blue rimmed dots). The gray dot corresponds to the energy crossing shown in (b) of those same states. (c) depicts the Maxwell construction and gives the coexistence points.}
    \label{fig:3d_loc_final}
\end{figure}

To explore whether such scenarios are also possible in the three-dimensional systems, we perform corresponding complex time evolutions for the above presented dipolar BEC model and establish the existence of localized states. For that matter, we would like to contextualize the state shown in Fig.~\ref{fig:3D_loc}, see Fig.~\ref{fig:3d_loc_final}. Again, the points of thermodynamic coexistence ($\bar{n}_\text{coex periodic}= 5237/\mu$m, $\bar{n}_\text{coex superfluid}= 5259/\mu$m) limit the interval where localized states can exist. Within this small interval, we are able to identify a localized state~\footnote{Here, for computational reasons we chose parameters that result in a first-order phase-transition independently of the domain size.}.

To conclude, we have shown that localized states exist in dipolar BECs. This raises a number of interesting points when exploring the thermodynamic limit and the order of the superfluid-supersolid phase transition. One implication is that one cannot decide whether a specific phase transition is of second order based on the occurrence of a supercritical bifurcation in a state diagram employing the mean density as control parameter as particles can be redistributed between coexisting states. One also has to explicitly exclude the possibility of thermodynamic coexistence of superfluid and supersolid states. Possible coexistence is evidenced by the existence of localized states as ground states that can be  established without the need for a complete (possibly computationally prohibitive) bifurcation analysis.

From the perspective of superfluidity and supersolidity localized states are interesting, as the wave function can feature a continuous transition from a perfect superfluid to a supersolid and ultimately possibly to insulating droplets in the same single realization. 

As an outlook, a comprehensive bifurcation analysis should be devoted to the three-dimensional case. In particular, it will be interesting to explore minimal finite BECs -- i.e., trapped in the $z$-direction as well -- for realistic experimental parameters (limited particle number and sufficiently low density) that can support localized states. Furthermore, it would be interesting to add temperature fluctuations and explore their effect on localized states in dipolar BECs~\cite{Sanchez-Baena:NatCom:2023,Baena:PRR:2024} and localized states in molecular BECs and their experimental feasibility~\cite{Bigagli2024}.

\begin{acknowledgments}
We acknowledge that parts of the calculations were performed on the HPC cluster PALMA II of the University of Münster, subsidized by the DFG (INST 211/667-1).
\end{acknowledgments}

%\bibliography{apssamp}
\input{SMGT2024-v04.bbl}
\end{document}